\documentclass[11pt,a4paper]{article} 
\usepackage{jheppub} 
\usepackage{amsmath}
\usepackage{graphicx}

\title{Gravitational collapse of k-essence}

\author[a]{Ratindranath Akhoury,}
\author[a,b]{David Garfinkle}
\author[a]{and Ryo Saotome}

\affiliation[a]{Michigan Center for Theoretical Physics, Randall Laboratory of Physics, University of Michigan, Ann Arbor, MI 48109-1120, USA}
\affiliation[b]{Department of Physics, Oakland University, Rochester, MI 48309, USA}

\emailAdd{akhoury@umich.edu}
\emailAdd{garfinkl@oakland.edu}
\emailAdd{rsaotome@umich.edu}

\abstract{We perform numerical simulations of the gravitational collapse of a k-essence scalar field.  When the field is sufficiently strongly 
gravitating, a black hole forms.  However, the black hole has two horizons: a light horizon (the ordinary black 
hole horizon) and a sound horizon that traps k-essence.  In certain cases the k-essence signals can travel
faster than light and the sound horizon is inside the light horizon.  Under those circumstances, k-essence signals
can escape from the black hole.  Eventually, the two horizons merge and the k-essence signals can no longer escape.}


\begin{document}
\maketitle
\flushbottom
\section{Introduction}

It has been experimentally verified that the universe has recently entered an era of accelerated expansion. This discovery has led physicists to pose the coincidence problem: why does the accelerated expansion of the universe coincide with the onset of matter domination? k-essence \cite{steinhardt,steinhardt2} has been proposed in order to answer this question. k-essence models postulate the existence of a scalar field with non-standard kinetic terms that allows for the field to have negative pressure once the matter dominated era begins.

It has been shown that if k-essence is to solve the coincidence problem, then fluctuations of the k-essence field must necessarily propagate superluminally at some point \cite{durrer}. However, it has been argued that as long as causality is not violated this should be physically acceptable \cite{vikman2}.

Given the possibility of superluminal propagation, it is interesting to ask how a k-essence field would act during gravitational collapse. It has been found that the stationary solution describing k-essence scalar field accretion into a background Schwarzschild black hole allows for the existence of two horizons, one corresponding to the fluctuations of the k-essence scalar field and the other being the normal light horizon \cite{vikman}. This is interesting as the existence of two horizons also occurs in other theories that allow superluminal motion \cite{ted,tedandme}.
It is possible that the stationary solution allowing for the existence of two horizons may be avoided kinematically given reasonable initial conditions. It is also possible that once back reactions of the scalar field on the metric are considered the two horizons may reduce to one. Another possibility is that gravitational collapse results not  in the formation of a black hole but a naked singularity.

In order to see which of the above mentioned possibilities are realized during the gravitational collapse of a k-essence scalar field, one can perform numerical simulations of this process. Such simulations have often been done before for an ordinary scalar 
field \cite{choptuik,gabor} and the methods used can be modified and generalized to treat k-essence. 
In this particular study, we will numerically simulate the self-consistent gravitational collapse of spherically symmetric scalar fields described by several different k-essence Lagrangians. The knowledge of what k-essence models allow for what type of horizons may help restrict the parameter space of acceptable k-essence models. Numerical simulations may also indicate whether or not there are any unique experimental signatures associated with the creation of a black hole from a k-essence scalar field.

The equations and numerical methods used are described in section 2 and the appendix.  
Results are presented in section 3 and conclusions discussed in section 4.  

\section{Methods}

The action for self gravitating k-essence is 
\begin{equation}
I = \int {\sqrt g} \, {d^4} x \, \left ( {\frac {{{}^{(4)}}R} {2 \kappa}} + {\cal L}(X) \right )
\end{equation}
Here ${{}^{(4)}}R$ is the spacetime scalar curvature, $\kappa = 8 \pi G$ where $G$ is Newton's gravitational constant, and 
$ X = - {\frac 1 2} {\nabla ^a} \phi {\nabla _a} \phi$ where $\phi$ is the k-essence scalar field.  Varying the action
with respect to the metric yields the Einstein field equation
\begin{equation}
{G_{ab}} = \kappa {T_{ab}}
\label{efe}
\end{equation}
where $G_{ab}$ is the Einstein tensor, and the k-essence stress-energy tensor $T_{ab}$ is given by
\begin{equation}
{T_{ab}} =  {{\cal L}_X}  {\nabla _a}\phi {\nabla _b} \phi + {\cal L} {g_{ab}}
\end{equation}
Here, we introduce the shorthand ${\cal L}_X$ to mean $d{\cal L}/dX$ (and later we will also use
${\cal L}_{XX}$ to mean ${d^2}{\cal L}/d{X^2}$).
Varying the action with respect to the k-essence scalar field $\phi$ yields the k-essence equation of motion
\begin{equation}
{{\tilde g}^{ab}} {\nabla _a}{\nabla _b} \phi = 0
\label{kefe}
\end{equation}
Here ${\tilde g}^{ab}$ is an effective inverse metric associated with the k-essence scalar field and is given by
\begin{equation}
{{\tilde g}^{ab}} = {{\cal L}_X}  {g^{ab}} - {{\cal L}_{XX}} {\nabla ^a}\phi {\nabla ^b}\phi
\label{geff}
\end{equation}

We perform numerical simulations of equations (\ref{efe}) and (\ref{kefe}) in the case of spherical symmetry.  For an 
ordinary scalar field, such simulations are often performed using polar-radial coordinates \cite{choptuik}.  Here the radial 
coordinate is chosen to be the usual area radius, while the time coordinate is chosen to be orthogonal to the radial coordinate.
However, polar-radial coordinates only last until the first trapped surface forms.  This would not be suitable for a simulation of 
k-essence gravitational collapse because k-essence travels faster than light and thus is not trapped by ordinary trapped surfaces
but instead by the trapped surfaces of the effective metric of eq. (\ref{geff}).  Thus a k-essence black hole has not yet formed 
when the first trapped surface forms and a simulation using polar-radial coordinates stops.  Instead, we use a different method
with a set of coordinates that includes a region within the black hole horizon.  This method uses maximal slicing, and a 
radial coordinate based on length rather than area, and is described in detail in the appendix.  

Another complication of k-essence collapse has to do with the surfaces of constant time.  In an evolution of the Einstein field 
equations, the surfaces of constant time must be spacelike.  That is, we must have ${g^{tt}} < 0$.  For an ordinary scalar field,
this condition is also enough to render the evolution of the scalar field well posed.  However, k-essence evolves in an
effective metric whose light cones are different from those of the spacetime metric.  In order that the surfaces of constant time
are also Cauchy surfaces for the k-essence field, we must have ${{\tilde g}^{tt}} < 0$.  In the simulations, it is sometimes
the case that ${{\tilde g}^{tt}} \to 0$ without any singularity forming and sometimes even without the formation of any trapped
surface.  In this case we are forced to stop the simulation at that point without obtaining a clear answer as to the outcome of the
collapse process.  This also suggests the disturbing possibility that the light cones of the k-essence field could tip so much that
no surface could be timelike for both the metric and the k-essence field.  In that case, the theory itself would cease to be well 
defined.  

In any case, our results are obtained by choosing a variety of k-essence Lagrangians 
$\cal L$ and a variety of initial data, and then evolving with
the maximal slicing method, and noting the outome of the collapse process.

\section{Results}

\begin{figure}
\includegraphics[width=3.0in]{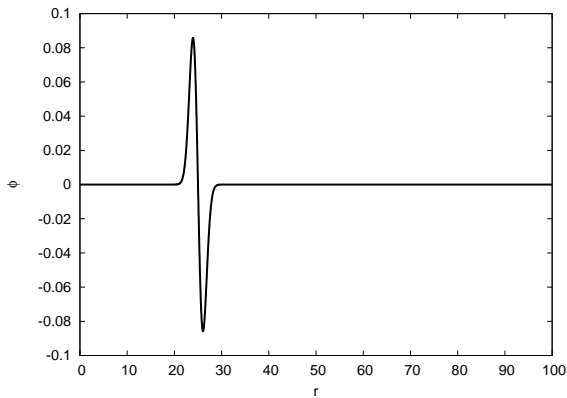}
\caption{\label{fig1}$\phi$ {\it vs} $r$ at the initial time $t=0$ for $A=0.15$}
\end{figure}

We now present the results of the collapse simulations.
We chose the scalar field to initially have ${\partial _t}\phi =0$ and the form:
\begin{equation}
\phi=-2 A \frac{r-r_{0}}{\sigma^{2}}\exp(\frac{-(r-r_{0})^{2}}{\sigma^{2}})
\label{initialphi}
\end{equation}
where $A$, $\sigma$, and $r_{0}$ are constants. For all results and figures shown we used the values $\sigma=1.5$, and $r_{0}=25$ (see fig \ref{fig1} for which $A=0.15$).
\subsection{Sound horizon Lagrangian}
We first consider a k-essence model proposed in \cite{vikman} with a Lagrangian density given by:
\begin{equation}
\mathcal{L}(X)=\alpha^{2}\left[\sqrt{1+\frac{2X}{\alpha^{2}}}-1\right]-\Lambda
\label{ryo1}
\end{equation}
We chose $\Lambda=0$ for all of our simulations. In \cite{vikman} it was found that when there is a steady flux of  scalar field described by this Lagrangian falling into a Schwarszschild black hole, there exists a stationary solution which can allow scalar field signals to escape from the light horizon. This is because the k-essence scalar field can propagate at speeds different from that of the speed of light. Thus there is a sound horizon associated with the scalar field which can differ from the traditional light horizon. When this sound horizon is located within the light horizon, signals can escape the black hole.

Before presenting the results, it is helpful to recall how this sort of initial data would evolve for an ordinary (non k-essence) 
massless free scalar field, both without and with gravity.  For an ordinary non-gravitating scalar field, 
initial data of the type given in
eq. (\ref{initialphi}) separates into an outgoing pulse and an ingoing pulse.  The ingoing pulse undergoes interference in a region 
near $r=0$ and becomes an outgoing pulse, so that at sufficiently late times the scalar field consists of two outgoing pulses.  For
an ordinary gravitating scalar field, when the amplitude $A$ is sufficiently small, the self-gravity of the field is weak, and the 
behavior is essentially that of a non-gravitating scalar field.  However, when $A$ is sufficiently large, as the ingoing pulse nears 
the center, it becomes trapped by its own self-gravity and forms a black hole.

For the model of eq. (\ref{ryo1}) when $\alpha^{2}$ is large compared to $X$, the Lagrangian approaches that of a massless free scalar field.  Therefore we would expect that for large enough $\alpha$ we would get the same sort of behavior as for an ordinary
(non k-essence) scalar field.  To see whether this expectation is right, we performed simulations with $\alpha =1.0$ and with
$A=0.05$ and $A=0.15$.  The result of the simulation with $A=0.05$ is shown in figure (\ref{fig2}).  Here the scalar field $\phi$ 
is graphed as a function of $r$ at time $t=45$.  In this amount of time the initially outgoing pulse has progressed to steadily larger $r$ with approximately $1/r$ falloff in amplitude, while the initially ingoing pulse has undergone interference near the center and become an outgoing pulse which has moved away from the center undergoing its own approximately $1/r$ falloff in amplitude.  

In contrast, figure (\ref{fig3}) shows the result of the simulation with $A=0.15$ (and $\alpha =1.0$).  Again the scalar field $\phi$ 
is graphed as a function of $r$ at time $t=45$.  Once again, the outgoing pulse has progressed to steadily larger $r$ with a corresponding falloff in amplitude.  However, now the ingoing pulse has not become outgoing and instead remains trapped near the
center by its own self-gravity.  Indeed we would expect the ingoing pulse to not merely ``remain trapped'' but rather to undergo complete gravitational collapse and form a singularity at the center of a black hole.  The fact that it has not done so in this simulation
is an artifact (and one of the main features) of the maximal slicing method.  In maximal slicing, the lapse function $N$ becomes small in regions that are strongly gravitating.  This slows the evolution down in those regions and prevents the simulation from reaching the singularity (which would force the simulation to halt).  Indeed such a ``collapse of the lapse'' is usually taken as an indication of
black hole formation.  This phenomenon is illustrated in figure (\ref{fig4}) where the lapse for this simulation is plotted.

Nonetheless, we would like a more precise indication of black hole formation.  In general relativity, such an indication is 
provided by the formation of a marginally outer trapped surface.  Here a two dimensional surface has an outgoing null geodesic normal
vector $\ell ^a$ and the surface is marginally outer trapped if ${\nabla _a}{\ell ^a} =0$.  This condition simplifies considerably in
the case of spherical symmetry, where it becomes equivalent to ${\ell ^a}{\nabla _a}R =0$ where $R$ is the area radius ({\it i.e.} $4 \pi {R^2} $ is the area of the spheres of symmetry).  That is, we want to know whether there is an outgoing vector $\ell ^a$
such that ${g_{ab}}{\ell ^a}{\ell ^b}=0$ and ${\ell ^a} {\nabla _a}R=0$.  For the coordinate system we use (eq. (\ref{metricform})) we find after some straight forward algebra that this condition becomes:
\begin{equation}
\frac{1}{2}RK^{r}_{r}+R'=0
\label{ryo2}
\end{equation}      
Thus, to check for the presence of a marginally outer trapped surface (and thus the formation of a black hole) we only need to compute
the left hand side of eq. (\ref{ryo2}) and see whether it vanishes somewhere. 

The condition of eq. (\ref{ryo2}) works for the gravitational collapse of an ordinary scalar field.  However, since k-essence can
travel faster than light, we want to know not only whether an object forms that can trap light, but also whether an object forms
that can trap k-essence.  Since k-essence satisfies a wave equation using the effective inverse 
metric ${\tilde g}^{ab}$, it follows that the direction of propagation of k-essence is a vector that is null with respect to
the effective metric ${\tilde g}^{-1}_{ab}$ which satisfies ${{\tilde g}^{-1}_{ab}}{{\tilde g}^{bc}}={{\delta ^c}_a}$.  
Straightforward calculation using eq. (\ref{geff}) then shows that  
\begin{equation}
{\tilde g}^{-1}_{ab}=\frac{1}{\mathcal{L}_{X}}g_{ab}+c_{s}^{2}\frac{\mathcal{L}_{XX}}{\mathcal{L}_{X}^{2}}\nabla_{a}\phi\nabla_{b}\phi
\end{equation}
where 
\begin{equation}
{c_s} = \sqrt{\frac {\mathcal{L}_{X}} {{\mathcal{L}_{X}} + 2 X {\mathcal{L}_{XX}}}}
\end{equation}
is the speed of propagation of the k-essence waves for this model.  (we will also refer to $c_s$ as the sound speed).  Thus
the condition for a sound horizon is just that for a light horizon, but with the effective metric replacing the metric.
That is, the existence of an outgoing vector $\ell ^a$ satisfying $ {{\tilde g}^{-1}_{ab}}{\ell ^a}{\ell ^b} =0$ and 
${\ell ^a}{\nabla _a}R=0$.  Some straightforward algebra using eq. ({\ref{metricform}) then shows that this condition is  
\begin{eqnarray}
\frac{1}{2}R{{K^r}_r}(\mathcal{L}_{X}+c_{s}^{2}\mathcal{L}_{XX}S^{2})
\nonumber
\\
+{R'}(\sqrt{\mathcal{L}_{X}(\mathcal{L}_{X}-2Xc_{s}^{2}\mathcal{L}_{XX})}-c_{s}^{2}\mathcal{L}_{XX}PS)=0
\label{ryo3}
\end{eqnarray}
In summary: the condition for a light horizon is the vanishing of the left hand side of eq. (\ref{ryo2}) while for a sound horizon
it is the vanishing of the left hand side of eq. (\ref{ryo3}).

\begin{figure}
\includegraphics[width=3.0in]{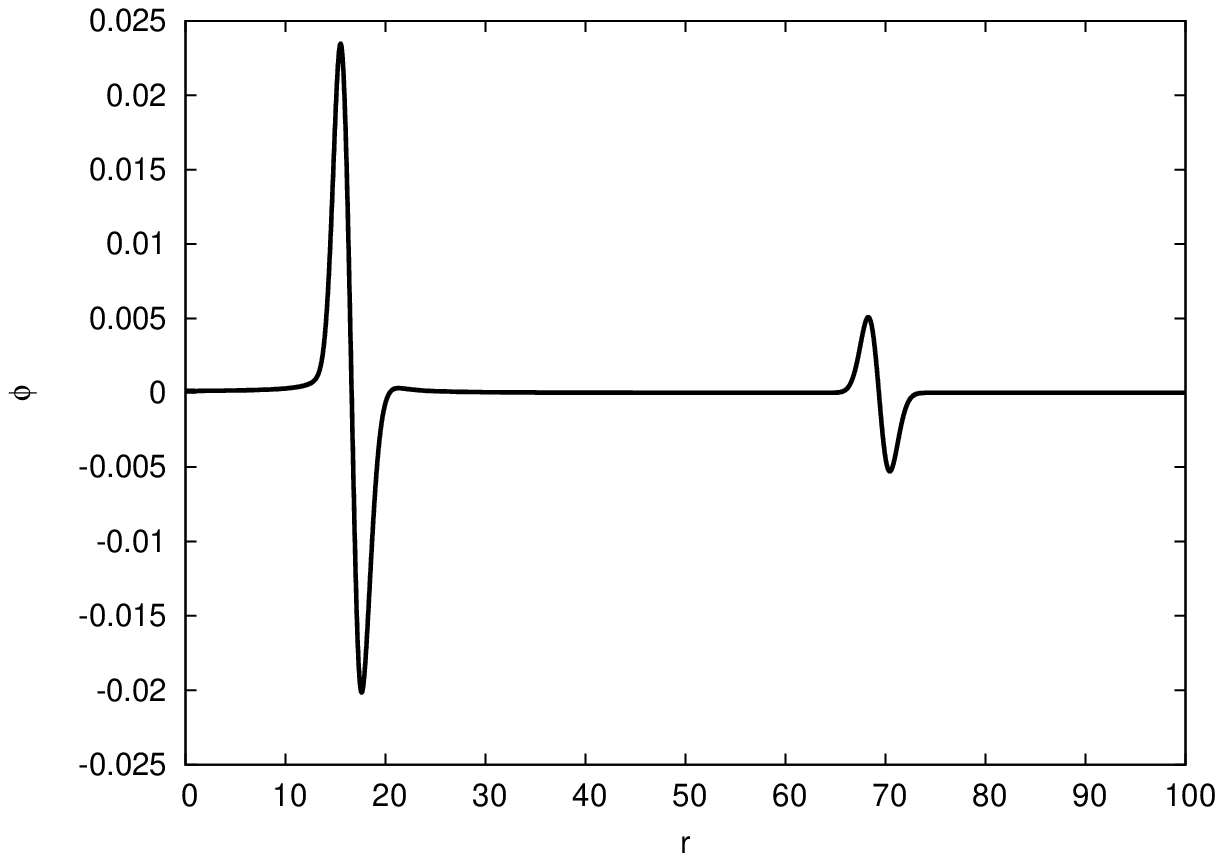}
\caption{\label{fig2}$\phi$ {\it vs} $r$ at time $t=45$ for $A=0.05$ and $\alpha =1.0$}
\end{figure}

\begin{figure}
\includegraphics[width=3.0in]{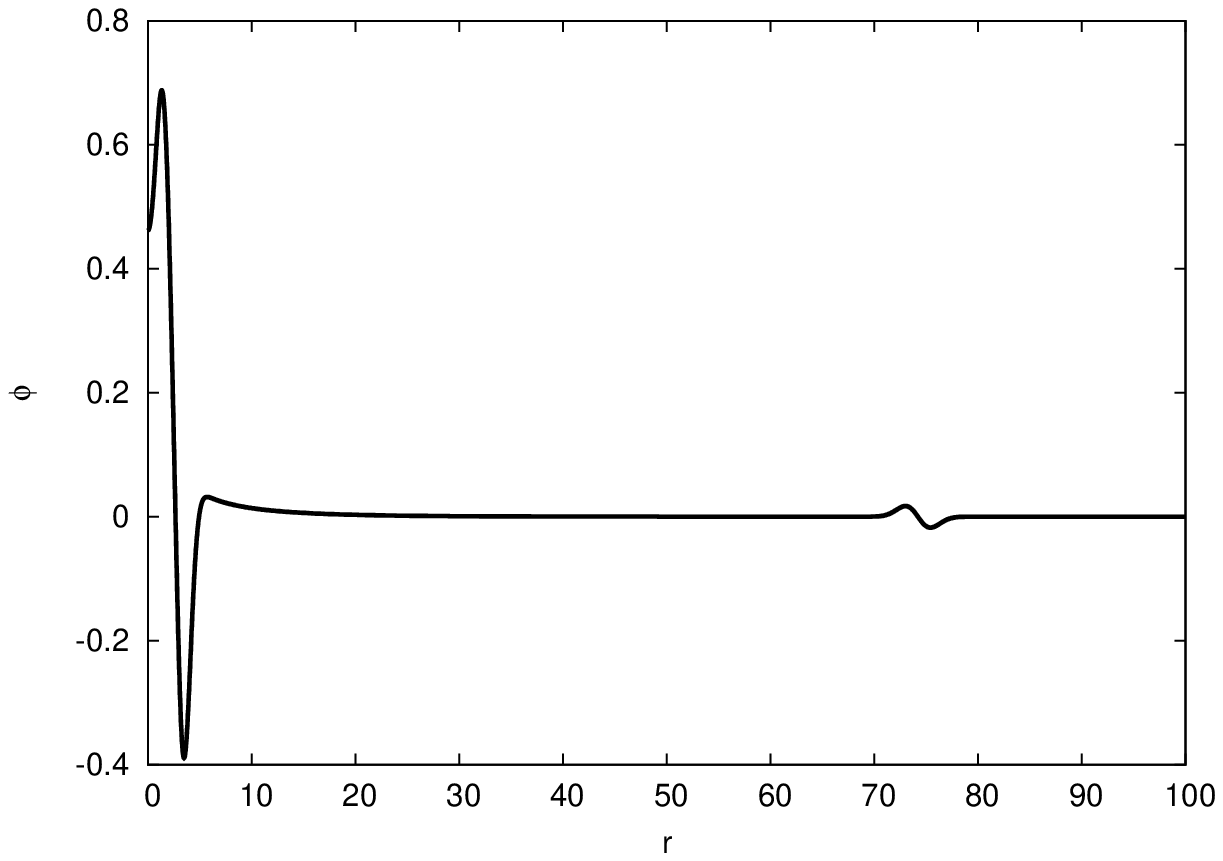}
\caption{\label{fig3}$\phi$ {\it vs} $r$ at time $t=45$ for $A=0.15$ and $\alpha =1.0$}
\end{figure}

\begin{figure}
\includegraphics[width=3.0in]{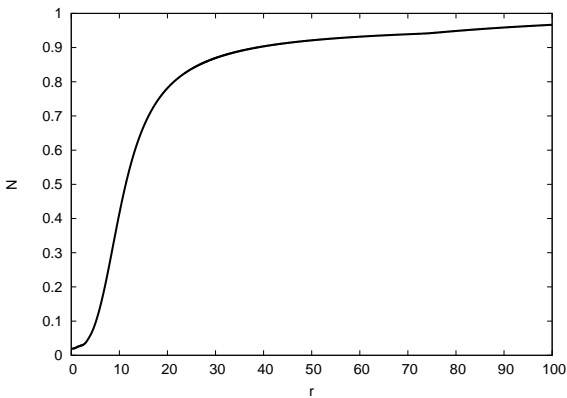}
\caption{\label{fig4}the lapse function $N$ {\it vs} $r$ at time $t=45$ for $A=0.15$ and $\alpha =1.0$}
\end{figure}

\begin{figure}
\includegraphics[width=3.0in]{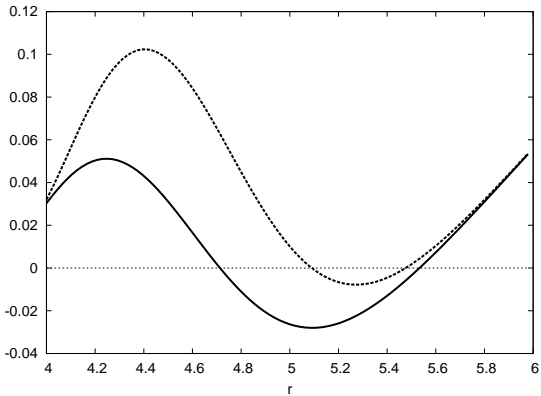}
\caption{\label{fig5}the conditions for light horizon (solid line) and sound horizon (dotted line) at time $t=38.8$ for $A=0.15$ and $\alpha =1.0$ and $\cal L$ given by eq. (\ref{ryo1})}
\end{figure}

In figure (\ref{fig5}) are plotted the left hand sides of eq. (\ref{ryo2}) (light horizon, solid line) and eq. (\ref{ryo3}) (sound horizon, dotted line) at a time shortly after the sound horizon first forms ($t=38.8$).  Note that each curve crosses zero more than once; however
it is the outermost crossing for each curve that is the horizon.  Also note that the sound horizon is inside the light horizon, as we would expect for k-essence that travels faster than light.  A similar comparison done at $t=45$ shows that both horizons have 
merged.  In physical terms, the reason for this merger is that eventually all scalar field either falls into the black hole or escapes
from it.  Thus at sufficiently late times there is no scalar field on the horizon, which thus makes the k-essence effective 
metric equal to the spacetime metric in that region.  Thus, the two horizons very quickly coincide early in the collapse process. This would indicate that it is only possible to send signals from inside the black hole for a very short time window right after a scalar wave packet falls into the black hole.  In contrast, the reason for the persistence of two distinct horizons in the stationary 
solution of \cite{vikman} is that the separation is maintained by a constant flux of scalar field through the horizon.

Since the Lagrangian of eq. (\ref{ryo1}) essentially coincides with that of a free scalar field for large $\alpha$ one might expect
to find behavior more uniquely characteristic of k-essence by performing simulations with small $\alpha$.  However, such simulations quickly run into difficulties.  In particular, a simulation with $A=0.15$ and $\alpha =0.5$ fails shortly after $t=34$.  At first sight this is somewhat puzzling, since at times shortly before the simulation fails, the scalar field shows no signs of the sort of rapid growth that would lead to singular behavior.  However, this mystery was quickly resolved by an examination of 
eq. (\ref{dtP}).  Note, that this equation involves division by the quantity ${{\cal L}_X} + {{\cal L}_{XX}}{P^2}$ and an 
examination of the simulation showed that this quantity was tending towards zero at a particular spatial point just as the simulation was about to fail.  One can understand this phenomenon geometrically as follows: from eqs. (\ref{geff}),
(\ref{metricform}) and (\ref{PSdef})
it follows that          
\begin{equation}
{\tilde g}^{tt}=-\frac{1}{N^{2}}(\mathcal{L}_{X}+\mathcal{L}_{XX}P^{2})
\end{equation}
So, it is evident that the program crashes when ${\tilde g}^{tt}=0$.  But this is precisely where the coordinate $t$ fails to
be timelike with respect to the k-essence effective metric, and therefore where the constant $t$ surfaces fail to be Cauchy surfaces 
for the k-essence field equation.  This sort of condition is familiar from general relativity, where the coordinates must be
chosen so that ${g^{tt}} < 0$ in order that the constant $t$ surfaces remain good Cauchy surfaces for the evolution of the 
spacetime metric.  However, when one is evolving self-gravitating k-essence it is necessary that the constant $t$ surfaces be timelike
with respect to both the spacetime metric $g^{ab}$ and the k-essence effective metric ${\tilde g}^{ab}$.   

In order to avoid this problem, one is led to consider Lagrangians other than that of eq. (\ref{ryo1}).  In particular, one could 
simply change the sign of coefficient of $\alpha ^2$ in eq. (\ref{ryo1}) to obtain
\begin{equation}
\mathcal{L}(X)=\alpha^{2}\left[1-\sqrt{1-\frac{2X}{\alpha^{2}}}\right]
\label{imalpha}
\end{equation}  

\begin{figure}
\includegraphics[width=3.0in]{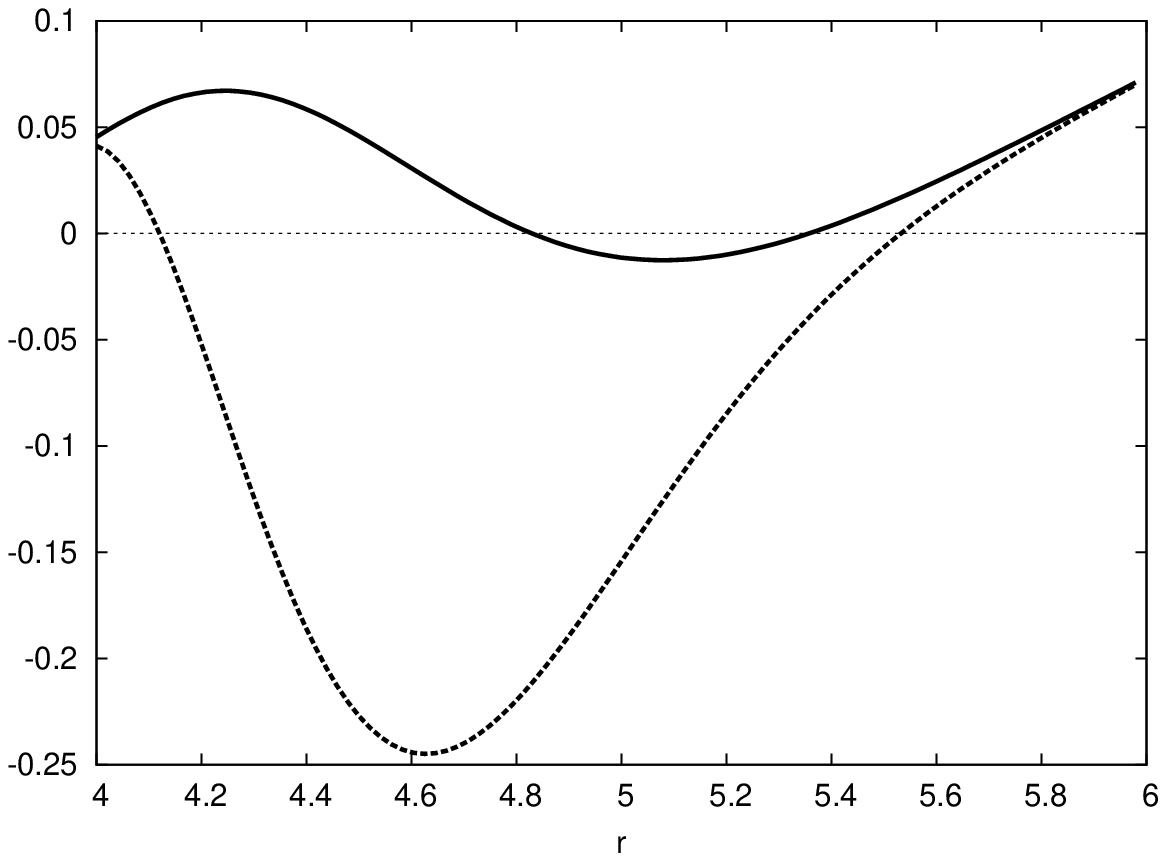}
\caption{\label{fig6}the conditions for light horizon (solid line) and sound horizon (dotted line) at time $t=38.5$ for $A=0.15$ and $\alpha =0.5$ and $\cal L$ given by eq. (\ref{imalpha})}
\end{figure}

In figure (\ref{fig6}) are plotted the results of a simulation using the Lagrangian of eq. (\ref{imalpha}).  Here $A=0.15$ and
$\alpha =0.5$ and the plot is done at a time ($t=38.5$) shortly after the formation of the light horizon.  As in figure (\ref{fig5})
what are plotted are the left hand sides of eq. (\ref{ryo2}) (light horizon, solid line) and eq. (\ref{ryo3}) (sound horizon, dotted line).  As before, the horizons occur at the outermost point where each curve crosses zero.  However, in this case the sound horizon is outside the light horizon, so there is no possibility of signals escaping from
the black hole.  Simulations using the Lagrangian of eq. (\ref{imalpha}) do not seem to have the problem with ${\tilde g}^{tt}$
vanishing.  Nonetheless, these simulations also fail at sufficiently small $\alpha$.  For example, a simulation with $A=0.15$ and
$\alpha =0.1$ fails at around $t=35$.  Here the problem is not ${\tilde g}^{tt}$ which remains negative, but rather the vanishing
of $c_s$.  This is a familiar phenomenon from simulations of hydrodynamics.  When the speed of propagation goes to zero, shock waves
tend to form.  One can still simulate such systems, but then special ``shock capturing'' numerical methods are needed rather than the 
simple finite differencing methods we use here.  

We also performed simulations of the Lagrangian given in eq. (\ref{ryo1}) in the absence of gravity. In this case, we found that for sufficiently small values of $\alpha$ the simulation still eventually fails due to ${\tilde g}^{tt}$ going to zero. This indicates that even the Minkowski time coordinate eventually ceases to be a valid global time coordinate for the emergent k-essence metric. As one might expect, no sound horizon was formed in this case.

On the other hand, it is possible to create a sound horizon in the absence of gravity if one simulates eq. (\ref{imalpha}) with a sufficiently small $\alpha$. As in the case where we did consider gravity, these simulations failed due to $c_{s}$ going to zero as opposed to ${\tilde g}^{tt}$ going to zero.

\subsection{Lagrangians where t is a valid global time coordinate}

Due to the problems encountered using the Lagrangians of eqs. (\ref{ryo1}) and ({\ref{imalpha}) we are led to consider Lagrangians with non-standard kinetic terms that do not allow either ${\tilde g}^{tt}$ nor $c_{s}$ to vanish. One such Lagrangian is:
\begin{equation}
\mathcal{L}=\frac{CX}{1+C}+e^{\frac{X}{1+C}}-1
\label{ryo4}
\end{equation}
where $C$ is a constant. Like the previous Lagrangians considered, this one has the desirable feature that it reduces to the free scalar field Lagrangian at small values of $X$. Note that as long as $C$ is positive, both $\mathcal{L}_{X}$ and $\mathcal{L}_{XX}$ are positive so ${\tilde g}^{tt}$ will not go to zero. Another requirement is that the k-essence wave equation (eq. (\ref{kefe}))
be hyperbolic. The condition for hyperbolicity of eq. (\ref{kefe}) is that ${\mathcal{L}_{X}}+2X{\mathcal{L}_{XX}}>0$. 
(this condition is satisfied for the Lagrangians of eqs. (\ref{ryo1}) and (\ref{imalpha})).  For the Lagrangian
of eq. (\ref{ryo4}), ${\mathcal{L}_{X}}+2X{\mathcal{L}_{XX}}$ has one global minimum with the value $\frac{C}{1+C}-\frac{2}{(1+C)e^{3/2}}$. Consequently, as long as we choose $C>0.45$, the k-essence wave equation will be hyperbolic.  The results of a simulation using this Lagrangian are plotted in figure (\ref{fig7}).  Here $A=0.15$ and $C=1$ and the plot is done at a time ($t=38.5$) shortly 
after the formation of the light horizon.  The plot shows the left hand sides of eq. (\ref{ryo2}) (light horizon, solid line) and eq. (\ref{ryo3}) (sound horizon, dotted line); and the horizons occur at the outermost point where each curve crosses zero.  Here, the 
sound horizon is slightly outside the light horizon, so no signal can escape from the black hole. 

\begin{figure}
\includegraphics[width=3.0in]{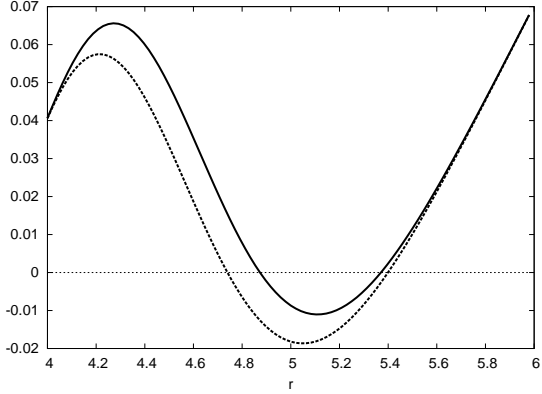}
\caption{\label{fig7}the conditions for light horizon (solid line) and sound horizon (dotted line) at time $t=38.5$ for $A=0.15$ and $C=1.0$ and $\cal L$ given by eq. (\ref{ryo4})}
\end{figure}

Another Lagrangian that doesn't result in either ${\tilde g}^{tt}$ or $c_{s}$ vanishing is
\begin{eqnarray}
&\mathcal{L}=X+CX^{2} &\mbox{ for } X > 0
\nonumber \\
&\mathcal{L}=X &\mbox{ for } X < 0
\label{ryo7}
\end{eqnarray}
It is clear that as long as $C$ and $X$ are positive, ${{\tilde g}^{tt}} < 0$ and ${\mathcal{L}_{X}}+2X{\mathcal{L}_{XX}}>0$. For negative $X$ the Lagrangian is that of a free field so there are no issues in that regime. 
The results of a simulation using this Lagrangian are plotted in figure (\ref{fig8}).  
Here $A=0.15$ and $C=1$ and the plot is done at a time ($t=38.5$) shortly 
after the formation of the light horizon.  The plot shows the left hand sides of eq. (\ref{ryo2}) (light horizon, solid line) and eq. (\ref{ryo3}) (sound horizon, dotted line); and the horizons occur at the outermost point where each curve crosses zero.  Note that there is a discontinuity in the dotted line.  This is due to the discontinuity at $X=0$ of the quantity ${\cal L}_{XX}$.  Here, the 
sound horizon is outside the light horizon, so no signal can escape from the black hole. 
Whether or not the sound horizon being outside the light horizon is a generic feature of k-essence models in which $t$ remains a valid global time coordinate is yet to be answered.
\begin{figure}
\includegraphics[width=3.0in]{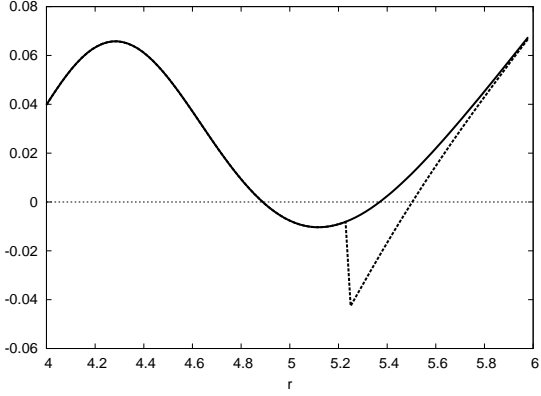}
\caption{\label{fig8}the conditions for light horizon (solid line) and sound horizon (dotted line) at time $t=38.5$ for $A=0.15$ and $C=1.0$ and $\cal L$ given by eq. (\ref{ryo7})}
\end{figure}

\subsection{Cosmological Lagrangians}

Other models of interest are cosmological k-essence models which serve as dynamical attractors that cause the field to serve as a cosmological constant at the onset of the matter dominated era. In general these Lagrangians are of a factorizable form in $\phi$ and $X$, that is they take the form: $\mathcal{L}=A(\phi)B(X)$. Two such examples that can be found in the literature are 
\cite{steinhardt,steinhardt2} :
\begin{equation}
\mathcal{L}(\phi,X)=\frac{1}{\phi^{2}}(-2.01+2\sqrt{1+X}+3 \cdot 10^{-17}X^{3}-10^{-24}X^{4})
\label{ryo5}
\end{equation}
and
\begin{equation}
\mathcal{L}(\phi,X)=\frac{1}{\phi^{2}}(-2.05+2\sqrt{1+f(X)})
\label{ryo6}
\end{equation}
where
\begin{eqnarray}
f(X)=X&-&10^{-8}X^{2}+10^{-12}X^{3}-10^{-16}X^{4}
\nonumber
\\
&+&10^{-20}X^{5}-10^{-24}X^{6}/2^{6}
\end{eqnarray}
Note during our simulations with these initial conditions, X does not get much larger than unity, so we can neglect terms that are higher order in X with small coefficients. Thus, for our purposes, both these Lagrangians reduce to approximately:
\begin{equation}
\mathcal{L}(\phi,X)=\frac{2}{\phi^{2}}(-1+\sqrt{1+X})
\end{equation}
The energy density of this field is proportional to $\frac{2}{\phi^{2}}$. As we know the universe is approximately flat, we require that any fluctuation of this field be a perturbation on a large constant field. Thus, the factor of $\frac{2}{\phi^{2}}$ in the Lagrangian will amount to a small overall factor. As the energy density will be negligibly small with large $\phi$, the gravitational effect will be small. Thus, both eq. (\ref{ryo5}) and eq. (\ref{ryo6}) should reduce to eq. (\ref{ryo1}) in flat space with $\alpha=1$ and a small overall rescaling factor multiplying the Lagrangian. This overall rescaling factor has no effect in the non-gravitational case. Simulations of eq. (\ref{ryo1}) in flat space with $\alpha=1$ result in no sound horizon but still suffer from having 
${\tilde g}^{tt}$ go to zero.

\section{Conclusions}

Our results show that the gravitational collapse of k-essence scalar fields can lead to black holes, just as it does for ordinary
scalar fields.  However, the nature of these black holes is different.  Rather than a single horizon they have two: a light horizon
and a sound horizon.  The light horizon is what we ordinarily think of as the black hole horizon.  The sound horizon (the place 
where the k-essence itself becomes trapped) can either be outside the light horizon (if the k-essence travels slower than light) or 
inside the light horizon (if the k-essence travels faster than light).  Thus, k-essence can allow signals to escape from a black hole. 
In the cases we study, the black hole is formed by the collapse of a pulse of k-essence.  Eventually all of this pulse either falls 
into the black hole or escapes from it.  Once that has happened, the two horizons have merged into one, and it is no longer possible for signals to escape from the black hole.  One should be able to maintain the two horizons, and the possibility of signals escaping
from the black hole, by feeding a black hole a steady stream of k-essence scalar field.  This question is currently under study \cite{agsv}.

None of our simulations give any indication of the formation of a naked singularity. However, in certain cases our simulations fail not because of any singular behavior in the spacetime itself, but rather because our constant time surfaces fail to be Cauchy surfaces for the evolution of the k-essence scalar field.  Since k-essence was originally proposed for cosmological models, one might wonder whether cosmological models might also be vulnerable to this sort of pathology.  Here, it is helpful to remember that cosmology treats a situation that is homogeneous and therefore where all physical quantities depend only on time.  For k-essence this means in particular that the quantity $X$ is always postive in cosmology. In our simulations of the k-essence model of eq. (\ref{ryo1}) the constant time surfaces can fail to be Cauchy surfaces for k-essence only
when $X$ is negative.  Thus, our gravitational collapse study brings to light certain properties of k-essence that would not be noticed in purely cosmological studies.  We have also proposed two k-essence Lagrangians for which this Cauchy surface problem does not occur.  It would be interesting to see if a Lagrangian of this sort could also be used for the usual purpose of k-essence: a natural cosmological explanation of dark energy. 

For those cases where the constant time surfaces fail to be Cauchy surfaces for k-essence, a natural solution to try would simply be to pick a different time coordinate.  However, for evolution of self-gravitating k-essence the constant time surfaces need to be Cauchy surfaces for {\it both} the gravitational field and the k-essence scalar field. It is not clear whether there is {\it any} time coordinate that will be able to do both jobs.  If there is no time coordinate that can do both jobs, then that 
particular k-essence Lagrangian would be incompatible with gravity and would therefore not be a physically sensible theory. However, our study yields no such assertion of incompatibility because we have used only one time coordinate: that of maximal slicing.  To resolve this issue, it is necessary to try other time coordinates to see whether they are compatible with both gravity and 
k-essence.  This question is currently under study \cite{mann}.         

\acknowledgments

We would like to thank Alex Vikman and Robert Mann for helpful discussions.  The work of DG was supported by NSF grant 
PHY-0855532 to Oakland University. RA and RS are supported by a grant from the US Department of Energy.

\appendix
\section{Maximal Slicing} 

Here we present the details of the maximal slicing method.  The metric is chosen to take the form
\begin{equation}
d {s^2} = - {N ^2} d{t^2} + {{(dr + {\beta ^r} dt)}^2} + {R^2} (d {\theta ^2} + {\sin ^2}\theta d {\varphi ^2})  
\label{metricform}
\end{equation}
which is the form for a spherically symmetric metric where the radial coordinate $r$ is length in the radial direction
in a constant time slice.  Note that the usual area radius $R$ is not one of the coordinates and is instead a function
of the coordinates $t$ and $r$.  Thus the spatial metric $\gamma _{ab}$ has components
\begin{eqnarray}
{\gamma _{rr}}=1
\nonumber
\\
{\gamma _{\theta \theta }} = {R^2}
\nonumber
\\
{\gamma _{\varphi \varphi}} = {R^2} {\sin ^2} \theta
\end{eqnarray}
The extrinsic curvature, $K_{ab}$ is defined by 
\begin{equation}
{K_{ab}} = - {{\gamma _a}^c}{\nabla _c}{n_b}
\label{extrinsic}
\end{equation}
where $n^a$ is the unit normal to the surfaces of constant time $t$.  However, due to spherical symmetry and maximal
slicing, there is only one independent component of the extrinsic curvature.  Specifically we have
\begin{equation}
{{K^\theta}_\theta}={{K^\varphi}_\varphi} = - {\textstyle {\frac 1 2}} {{K^r}_r}
\end{equation}
Equation (\ref{extrinsic}) is equivalent to 
\begin{equation}
{\partial _t} {\gamma _{ij}} = - 2 N {K_{ij}} + {D_i}{\beta _j} +{D_j}{\beta _i}
\label{gammadot}
\end{equation}
where $D_i$ is the covariant derivative of the spatial metric $\gamma _{ij}$.  The $rr$ component of 
eq. (\ref{gammadot}) yields
\begin{equation}
{\partial _r}{\beta ^r} = N {{K^r}_r}
\end{equation}
whose solution is 
\begin{equation}
{\beta ^r} = {\int _0 ^r} N {{K^r}_r} \; d r
\label{shift}
\end{equation}
The $\theta \theta $ component of eq. (\ref{gammadot}) yields
\begin{equation}
{\partial _t} R = {\beta ^r} {\partial _r} R + {\frac N 2} R {{K^r}_r}
\label{dtR}
\end{equation}

We now use the momentum constraint of the Einstein field equation to determine the extrinsic curvature.  For maximal slicing
($K=0$) this constraint is
\begin{equation}
{D_a}{K^{ab}} = - \kappa  {\gamma ^{bc}}{n^d}{T_{cd}}
\label{momentum}
\end{equation}
Define the quantities $P$ and $S$ by 
\begin{equation}
P = {n^a}{\nabla _a} \phi , \; \; \; S = {\partial _r} \phi
\label{PSdef}
\end{equation}
Then the quantity $X$ is given by 
\begin{equation}
X = {\textstyle {\frac 1 2}} ({P^2} - {S^2})
\label{Xcalc}
\end{equation}
while eq. (\ref{momentum}) becomes 
\begin{equation}
{R^{-3}} {\partial _r} ({R^3} {{K^r}_r}) = - \kappa {{\cal L}_X} P S
\label{momentum2}
\end{equation}
The solution of eq.(\ref{momentum2}) is 
\begin{equation}
{{K^r}_r} = - \kappa {R^{-3}} {\int _0 ^r} {R^3} {{\cal L}_X} P S \; dr
\label{Krr}
\end{equation}

Note that there is also a Hamiltonian constraint associated with the Einstein field equation.  In the case of maximal slicing, 
this constraint is 
\begin{equation}
{{}^{(3)}}R - {K_{ab}}{K^{ab}} = 2 \kappa {T_{ab}}{n^a}{n^b}
\end{equation} 
where ${{}^{(3)}}R$ is the spatial scalar curvature.  This equation yields
\begin{eqnarray}
{\partial _r}{\partial _r} R &=& {\frac {1 - {{({\partial _r}R)}^2}} {2R}} 
\nonumber
\\
&-& {\textstyle {\frac 1 4}} R \left [ 
{\textstyle {\frac 3 2}} {{({{K^r}_r})}^2} + 2 \kappa ( {{\cal L}_X}{P^2} - {\cal L})\right ]
\label{hamilton}
\end{eqnarray}

We now determine the lapse $N$.  It follows from the maximal slicing condition that 
\begin{equation}
{D_a}{D^a} N = N \left [ {K_{ab}}{K^{ab}} + {\frac \kappa 2} {T_{ab}}({n^a}{n^b}+{\gamma ^{ab}})\right ]
\end{equation}
which yields
\begin{eqnarray}
&{\partial _r}{\partial _r} N& + {\frac 2 R} ({\partial _r}R )({\partial _r}N )
\nonumber
\\
&=& N \left [ 
{\textstyle {\frac 3 2}} {{({{K^r}_r})}^2} + \kappa \left ( {\textstyle {\frac 1 2}} {{\cal L}_X} ( {P^2} +{S^2} ) +
{\cal L} \right ) \right ]
\label{lapse}
\end{eqnarray}

We now consider the evolution of the k-essence scalar field.  From the definitions of $P$ and $S$ it follows that
\begin{eqnarray}
{\partial _t} \phi = N P + {\beta ^r} S
\label{dtphi}
\\
{\partial _t} S = N ({\partial _r}P + {{K^r}_r} S) + P {\partial _r} N + {\beta ^r}{\partial _r} S
\label{dtS}
\end{eqnarray}
The k-essence equation of motion, eq. (\ref{kefe}) becomes after some straightforward but tedious algebra
\begin{eqnarray}
{\partial _t} P &=& {\beta ^r}{\partial _r} P + S {\partial _r} N
\nonumber 
\\
&+& {\frac {N {{\cal L}_{XX}}} {{{\cal L}_X} + {P^2}{{\cal L}_{XX}}}} \left ( {{K^r}_r} P {S^2} + 2 P S {\partial _r} P
- {S^2} {\partial _r} S \right ) 
\nonumber 
\\
&+& {\frac {N {{\cal L}_X}} {{{\cal L}_X} + {P^2}{{\cal L}_{XX}}}} {R^{-2}} {\partial _r} ({R^2} S) 
\label{dtP}
\end{eqnarray}

The initial data are chosen as follows: $P$ and $\phi$ are chosen freely, with $S$ set equal to ${\partial _r} \phi$.  
Then $X$ is determined using eq. (\ref{Xcalc}) and then used to find
${\cal L}$ and  ${{\cal L}_X}$  Then eq. (\ref{Krr}) is integrated to find ${K^r}_r$.  Then
eq. (\ref{hamilton}) is integrated for $R$ using the fact that at the origin $R=0$ and ${\partial _r}R=1$.  
Then at each time step, the evolution proceeds as follows: first $X$ is determined using eq. (\ref{Xcalc}) and then used to find
${\cal L}, {{\cal L}_X}$ and ${\cal L}_{XX}$.  Then eq. (\ref{Krr}) is integrated to find ${K^r}_r$.  
And then eq. (\ref{lapse}) is solved for the lapse $N$
(using a tridiagonal method and the fact that 
${\partial _r}N =0$ at the origin and $N \to 1$ at infinity).
Then eq. (\ref{shift})
is integrated to find the shift $\beta ^r$.  Finally, the quantities $R, \, \phi , \, S$ and $P$ are
evolved to the next time step using eqs. (\ref{dtR}), (\ref{dtphi}), (\ref{dtS}) and (\ref{dtP}) respectively.  The 
evolution is done using the iterated Crank-Nicholson method, and all spatial derivatives are found using standard centered 
differences.

\end{document}